\documentclass[jidm,a4paper]{jidm} 
\usepackage{graphicx,url}  
\usepackage[utf8]{inputenc}
\usepackage[T1]{fontenc}

\newdef{definition}[theorem]{Definition}
\newdef{remark}[theorem]{Remark}

\jidmVolume{3}
\jidmNumber{3}
\jidmYear{12}
\jidmMonth{October}
\title{Location-Based Events Detection on Micro-Blogs}
\author{Augusto Dias Pereira dos Santos, Leandro Krug Wives, Luis Otavio Alvares}

\institute{PPGC/UFRGS, Brazil \\ \email{\{adpsantos, wives, alvares\}@inf.ufrgs.br}}

\begin{abstract}
The increasing use of social networks generates enormous amounts of data that can be used for various types of analysis. Some of these data have temporal and geographical information, which can be used for comprehensive examination. In this paper, we propose a new method to analyze the massive volume of messages available in Twitter to identify places in the world where events such as TV shows, climate change, disasters, and sports are emerging. The proposed approach is based on a neural network used to detect outliers from a time series, which is built upon statistical data from tweets located in different political divisions (i.e., countries, cities). These outliers are used to identify localized events within an abnormal behavior in Twitter. The effectiveness of our method is evaluated in an online environment indicating new findings on modeling local people's behavior from different places.
\end{abstract}



\begin{document}

\begin{bottomstuff}
\end{bottomstuff}

\maketitle


\section{Introduction}

Modeling the human behavior has always been an attempt of several scientists, and with social networks this task can be done in many perspectives. Social networks allow people to interact on the Internet as they do in the real world, sharing their lives through text messages, photos, videos, and connecting to friends with comments, likes, quizzes and games. It is important to state that we follow the definition of \cite{Wellman1996} regarding social networks, who states that when computer networks link people as well as machines, they become social networks. Some social networks in particular focus on sharing users' short text messages. These are called micro blogs, since they are similar to web blogs but with just a few words, being very attractive to mobile appliances. The most popular micro blog is Twitter, and due to an easy-to-use API it is widely used in many mobile and desktop platforms. Twitter was launched in 2006, and after 6 years it has around 140 million active users sending an average of 340 million tweets, those short messages, per day\footnote{http://blog.twitter.com/2012/03/twitter-turns-six.html}. The public default policy of tweets enables researches of various areas to be done on subjects that may vary from natural language processing and data mining to public health analysis. We suggest reading the first quantitative study \cite{Kwak2010} on the entire Twitter and its information diffusion to better understand Twitter's topology, influential identification and trending topics' behavior.

Using Twitter in mobile devices makes it possible to embed geographical information in the tweets. Tweets stored within GPS coordinates or political division names enable us to identify from where these messages were sent and conduct a socio-geographic analysis. 

Socio-geographic data is very difficult to be obtained. Cellular service providers, vehicle GPS trackers and credit card companies are some examples of businesses that have these data, but lock them with strict security \cite{Ferrari2012}. Some academic researches even needed to build their own set of data to study some socio-geographic patterns \cite{Li2008} \cite{Lerin2011}.

This is why public data from social networks bring these researches to a new level with live, organic and enormous amount of data. That way, human behavior can be modeled, identifying what the users from a certain city or place are saying about a specific topic and why, i.e., what their impressions are. 

Along with its real-time nature, Twitter information can be used as a live sensor network, for instance, detecting earthquakes and typhoons \cite{Sakaki2010} or local social events \cite{Lee2010}. In this paper, a topic is some subject referred to in a document and which users are talking about at any particular time, and an event means a unique thing that happens at some point of time \cite{Allan1998} \cite{Allan2002}.

In this context, this paper proposes a new method for using the vast volume of Twitter's user messages to identify location-based events such as concerts, festivals, disasters, political demonstrations, etc., without having to select keywords. This points to our main contribution on event detection, changing the dimensional space from keywords to places. In this sense, the Twitter's Streaming API\footnote{https://dev.twitter.com/docs} method is used to retrieve geo-tagged and time-stamped short text messages at a worldwide coverage. Simple metrics are extracted from these messages, considering political divisions as partitions, creating time series and used as input of a neural network \cite{HEINEN2011} that models the input data based on a regression technique and identifies outliers. Text messages are then parsed to provide semantic information to the events detected. 

The paper is organized as follows: section \ref{sec:related} presents related works; in section \ref{sec:method}, we present the proposed approach for location-based event detection; section \ref{sec:experiments} illustrates the experimental results and more detail on how the approach solves this task; and section \ref{sec:conclusion} provides the conclusions and discussion of further works.

\section{Related Works}\label{sec:related}

This section presents and discusses related works in the fields of Geo-social analysis and event detection, which are the main applications of our work.

\subsection{Geo-Social Analysis}\label{sec:GSA}

Despite the early stage of location-based social networks, or social network with some location information, many researches are being conducted to extract some knowledge from geo-social relations, in order to improve the location prediction of individuals in a social network better than with IP-based geo-location. Backstrom et al. \cite{Backstrom2010} used user-supplied addresses and the network of relation between profiles of the Facebook social network. Besides performing 69.1\% of accuracy with their best method, against 57.2\% for IP location, some interesting geo-social relations were confirmed, as intuitively known: people living in metropolitan areas are more cosmopolitan; they are more likely to have ties to distant places; the higher the population density, the lower the probability of knowing a person inside a square mile; and, in their data, 96\% of people live in areas less dense than 50 people per square mile.

For geographic mood characteristics analysis, Mislove et al \cite{Mislove2010} analyzed tweets posted from September 2006 to August 2009, extracting words containing psychological rating, according to ANEW system \cite{Bradley1999}, and matching them with the user profile location to identify some mood variations over the week, the hours of the day and the costs of the United States. These messages suggest that the West coast is happier than the East coast, and that happiness peaks occur each Sunday morning, with a trough on Thursday evenings, having the early morning and late evening the highest level of happy tweets. These works model some aspects of human behavior, but using static geographical information. Our study focuses on using information that changes in time and space with greater rate.

Due to its real-time property and massive adoption in the world, Twitter can be used as a sensor network for natural and social event detection, sometimes before its coverage by the news media or the government.  In Sakaki's \cite{Sakaki2010} work, they use geo-located tweets that have keywords related to natural hazard events such as \textit{earthquake} or \textit{shaking} to detect such events. With particle filtering, they can estimate the centers of earthquakes and the trajectories of typhoons, detecting 96\% of earthquakes, with seismic intensity scale of 3 or more, registered by Japan's Meteorological Agency.

In a recent work, Lee \cite{Lee2010} developed a system to discover unusual regional social activities using Twitter geo-tagged information. Their framework has four steps: Collecting crowd experiences via Twitter, establishing natural socio-geographic regions, estimating geographical regularity of local crowd behavior, and detecting unusual geo-social events. The first step uses a divide and conquer solution to solve the Twitter Search API restriction of 1,500 results per query. The second uses K-Mean clustering algorithm with Voronoi's diagram \cite{kmeans} to create socio-geographic regions, a step that can impact a online system. On the third one, three metrics are estimated for each cluster, hourly: number of tweets, number of users, and movement of local crowd. The last step divides the day in 6-hour periods and calculates the regularities of each cluster's metric using box plots that can also detect unusual statuses.

This method detected 903 unusual activities from 7,200 possible (300 clusters x 6 days x 4 periods (6-h)) and compared to the investigated list of 50 events, from Japan's local event guide site, 32 of them could be found, resulting in a recall performance of 64\% (32/50) plus a precision rate of 3.54\% (32/903). We must consider that this list is somewhat restricted, because other unexpected events, off the list, occurred and were detected. Despite the great advances in local event detection, driven primarily by the movement of local crowd's metric, there are some deprecated issues, unnecessary steps and heavy processing.

\subsection{Event Detection}\label{sec:EDT}

Event detection and tracking is a subset of problems from topic detection and tracking (TDT). The early definitions are from \cite{Allan1998,Allan2002}, in an initiative to investigate the state-of-the-art on finding and following new events in a stream of broadcast news stories. With the huge amount of information available on-line, the World Wide Web is a fertile source for that kind of event detection, and web mining research is at the crossroad of research from several research communities \cite{Kosala2000}. Over the last 10 years, user-generated content has come to dominate a large portion of the web and a real-time web has arisen to challenge number of areas of research, notably information retrieval and web data mining \cite{Bermingham2010}.

Becker \cite{Becker2011} presents a task of event identification on Twitter that is based on text analysis and clustering approaches, and shows numerous categories of features that must be considered: temporal, social, topical, and Twitter-centric. He also analyzes the different features that can impact the performance of a real-time system for event detection. The proposed technique for event identification offers a significant improvement over other approaches, showing that they can identify real-world event content in a large-scale stream of Twitter data. The use of location-based signals in event identification is suggested for future work.

A filtered stream of tweets to automatically identify events of interest, using just the volume of tweets generated at any moment of an event, was suggested by \cite{Lanagan2011} to provide a very accurate means of event detection, as well as an automatic method for tagging events with representative words from the tweet stream. That approach leads to the problem of choosing a set of words and tags that represent a field of interest, missing any other event that doesn't match it.

\section{The proposed approach}\label{sec:method}

\begin{figure}[t]
\centering
\includegraphics[width=110mm]{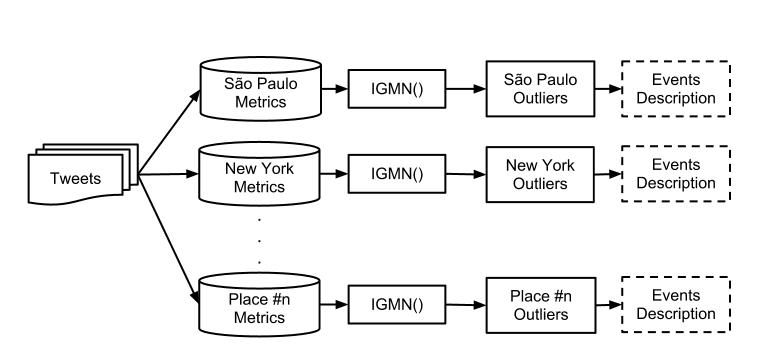}
\caption{Proposed data flow\label{figFramework}}
\end{figure}

To achieve the detection of events based on location using the huge amount of data provided by Twitter, we proceeded with the simpler data flow possible that lead to this goal. Figure \ref{figFramework} shows these flow as described below:

\begin{itemize}
\item Tweets: A crawler collects tweets from Twitter using Streaming API service;
\item Places Metrics: Creates two \textit{time series} from the number of tweets and users in a \textit{time instance} (or \textit{bin});
\item IGMN: The neural network is used to create data models and identify outliers;
\item Place Outliers: Consist in the time instances that were detected as outliers in both time series;
\item Events Description: Through the messages contained in the \textit{time instance} outliers it is possible to evaluate and understand the triggered event.
\end{itemize}

In relation to the crawler, it is important to state that Twitter's Streaming API is one of many Twitter's public services available. It allows real-time access to various subsets of public tweets with high throughput. Any message sent to the social network, with public permission and that matches a given query, will be delivered to the crawler. This service has filter parameters such as tracking some keyword occurrences in status messages, following tweets from a specific set of users or specifying a set of geographic bounding boxes to track. In this aspect, it is important to state that, since September 2010, the bounding box can be of worldwide coverage, allowing the retrieval of all tweets in a single query, and thus there is no need any more to build a monitor system as Lee \cite{Lee2010} suggests.

Each status message given by this API contains the text of the message, its creation's date/time, the message's id, the id and the full profile information of the user that has sent the message, and, sometimes, both place/country name and latitude/longitude, or just one of them. This happens because this information is sensible and for the sake of privacy the user may state whether or not he wants to share such specific latitude/longitude information or just the place's name. Current localization technology used by Twitter comprises GPS and GPS-A (which have latitude/longitude information) and originating IP (which has not latitude/longitude information).  The location technology used can also be retrieved, if allowed by the user, besides information given by the Twitter's geographic database (which doesn't have all world's countries, provinces/states, cities, neighbors and areas names).

For the last problem, we use a geographic database source\footnote{http://geocommons.com/overlays/85161} to translate those latitude/longitude information into names that are not known by Twitter . For instance, many Eastern countries and cities have blank names in the service API. So, this step is important because all our analysis is based on grouping tweets in sets of places as shown in Figure \ref{figFramework}. This location identification process is made during real-time streaming consumption.

Once the messages are localized (i.e., have location information), the next step consists on the identification of events. For this task, as stated before, we use a neural network (IGMN) to analyze time series and find outliers. A time series is a sequence of observations occurring in equal time intervals, having some basic properties/components \cite{Brockwell1986}. In a time series there are different components, for instance, seasonal component, trend component, and so on. The seasonal component describes when the time series' data experience regular changes which recur in some period of time (e.g., daily, weekly, monthly, and so on). The trend component indicates a series with upward or downward long term movement. Thus, the series is stationary when the mean, variance and autocorrelation structures do not change over time, and doesn't have a trend. A multivariate time series has more than one variable, while a univariate time series has only one variable. Our data can be described as a stationary, seasonal and univariate time series.

After the time series analysis is performed, we apply specific metrics to detect events. The metrics used in this work are extracted by grouping the text messages in sets of cities, provinces/states or countries, depending on the amount of information in each instance, then computing the number of users and number of tweets, creating two separate time series. We have chosen simple metrics like these because our intention was to develop a real time on-line event detection system. So we needed to decrease the framework's processing time. The usage of geographic names improved the framework in two ways: 

\begin{itemize}
\item Despite the linear complexity of K-Means, used on \cite{Lee2010}, there is no need to use clustering algorithms, since the message clustering is based on political divisions;
\item We increased the amount of analyzed tweets using all types of messages: 
    \begin{itemize}
        \item With and without GPS features; and/or
        \item With and without places' names.
    \end{itemize}
\end{itemize}

Once this splitting is done, we have a set of \textit{m} messages for each political division chosen. Thus the metrics are collected for each time instance (1 minute, 10 minutes, 1 hour, 6 hours, etc.) during a period of \textit{d} days, creating a time series. Lee's approach \cite{Lee2010} splits the day in 6-hour periods and uses box plot statistical analysis to detect outliers. We have discovered that this 6-hour period can hide some interesting detailed information about events happening in these political division areas, because the tilt's curve is relevant in a 6-hour slice, smoothing the mid curve outliers and empowering the beginning/end period data outliers. Beyond that, box plot is a univariate statistical tool \cite{Hardle2012} and the Twitter stream has a temporal dependency, as can be observed in Figure \ref{figOsloMunichSP}. The term univariate has different meaning in time series analysis, it refers to a time series that consists of single (scalar) observations recorded sequentially over equal time increments, time is in fact an implicit variable in the time series\footnote{http://www.itl.nist.gov/div898/handbook/pmc/section4/pmc44.htm}.

For the outliers detection task we use the Incremental Gaussian Mixture Network (IGMN) \cite{HEINEN2011}, a neural network that creates and continually adjusts probabilistic models consistent to all sequentially presented data, after each data point presentation, and without the need to store any past data points. Its learning process is aggressive, or "one-shot", meaning that only a single scan through the data is necessary in order to obtain a consistent model. Compared to (S)ARIMA \cite{Brockwell1986} has equivalent root mean square error without the need to pre-understand the time series components and data correlation imposed to (S)ARIMA's parameters, that facilitates the process of adding new places to the framework. The incremental process is another advantage against (S)ARIMA that needs a long period of data to model time series, that makes it possible to extend the framework for real-time analysis of Twitter stream.

After the outliers detection phase, each outlier represents a time instance that is analyzed for its content. Which event triggers these outliers? We collect all messages in this time instance. Those messages are processed in a search of most frequent words, ignoring stop words. The stop word database needs to be rebuilt to the short text message context, which uses a lot of abbreviations. These top rank words can provide us with a great idea of the triggered event, confirmed or not by the web and news search over the Internet.

\section{Experiments}\label{sec:experiments}

For performing our experiments we have collected data from Twitter since January 2011. We have adjusted the \textit{locations} parameter of the Twitter's Streaming API to the bounding box corresponding to (-179.99, -89.99, 179.99, 89.99), which relates to the entire globe.  Today we count with more than 1.4 billion geo-tagged messages, and around 10 million users. Considering this data set we have found that these users produced about 4.1 million geo-tagged tweets per day, where 42.25\% contained geographic coordinates, and 93.49\% contained places' names. 

With an on-line collecting system, a routine calculated countries' tweets of non-set country messages using the country boundary geographic database in a PostGIS server server\footnote{http://postgis.refractions.net/}. Data were stored in a MySQL\footnote{http://www.mysql.com/} database with a single structure: tweets' and users' tables, indexing \textit{message id}, \textit{user id}, \textit{created at} timestamp, \textit{country} and \textit{city} columns for faster grouping by clause. A 3-tier architecture provided more concurrence in order to avoid overload in the database; one server is the \textit{collector}, sending packages of 30 minutes' data to the \textit{data storage} computed by the \textit{processor} that generates the time series, detects the outliers and fetches the most frequent words used to describe the event.

The first step to create our time series is to choose a political division or place. We have five types of places, from Twitter definitions: country, admin (province/state), city, neighbor and POI (i.e., points of interest like restaurants, stores, museums, etc.); from wide to narrow areas. The wider the area, the more tweets per second are generated, but some places have a greater rate than others. Besides that, as more restrict is the area, the more local the event, we need a minimum number of messages per time instance in order to make the time series smoother. If we get few tweets per bin, the time series gets fluctuated values. The bin's size, which determines the amount of messages, needs to be evaluated to each place in order to identify which value gets the best event detection.

\begin{figure}[t]
\centering
\includegraphics[width=1\textwidth]{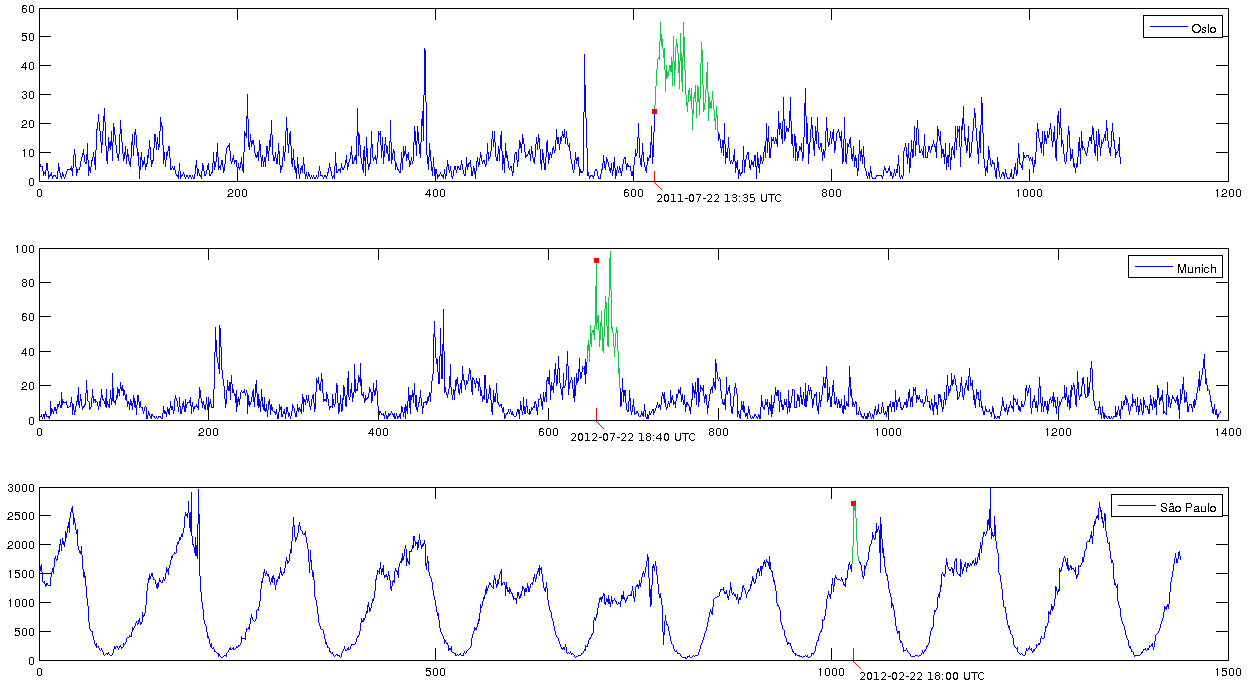}
\caption{Sample of events from Olso, Munich and São Paulo \label{figOsloMunichSP}}
\end{figure}

Figure \ref{figOsloMunichSP} shows some samples of tweets' time series generated with a bin of 10 minutes, for visualization purposes, in which it is easy to see a pattern of daily seasonality, represented by 144 values per day. Ordered by the volume of messages per bin, this figure shows events with different characteristics, all of them identified as outliers by our approach. The real date and time in which the event starts is indicated in the figure as its disturbance on the time series:

\begin{itemize}
\item Oslo bombing event: great disturbance on time series and long duration;
\item Munich soccer match: great disturbance and short duration;
\item São Paulo carnival vote counting: small disturbance and short duration
\end{itemize}

Once the bin size is chosen, two time series are made:
\begin{itemize}
\item Tweets time series (\textit{TweetsTS}): each value represents the amount of messages sent to Twitter server in one time instance;
\item Users time series (\textit{UsersTS}): each value represents the number of unique users who have sent messages to Twitter at that time instance.
\end{itemize}

To obtain the relevant outliers, each time series is modeled by the neural network, which returns the outliers of each one. An outlier is considered relevant when a time instance is detected as an outlier in both time series. It is noteworthy that the IGMN consider the values that are above or below the local likelihood as being outliers. However, in this work, we are only interested in the values above such likelihood, since they represent data beyond the normal volume. 

\begin{equation}\label{equation1}
Outliers = Intersect ( \textit{TweetsTS}.outliers\_above , \textit{UsersTS}.outliers\_above )
\end{equation}

Another parameter can be tuned to result in better quality events. The IGMN adjusts its models to the presented data using clustering techniques, and the similarity between the inputs is measured by the probability of each input belonging to the existing clusters. In this sense, the standard deviation may be used to indicate when a new cluster must be created, i.e., if the new data is too different from any cluster, this parameter is used to detect if a given input should be considered an outlier, based on the local likelihood.

For preliminary analysis and to evaluate the method's precision over different parameters, we have chosen the city São Paulo, Brazil, as a place (political division), because it is the number one city in the world in volume of tweets with geographic information. For this article the period from 2012-02-19 to 2012-02-24 was selected for those tests be done

\begin{figure}[t]
\centering
\includegraphics[width=1\textwidth]{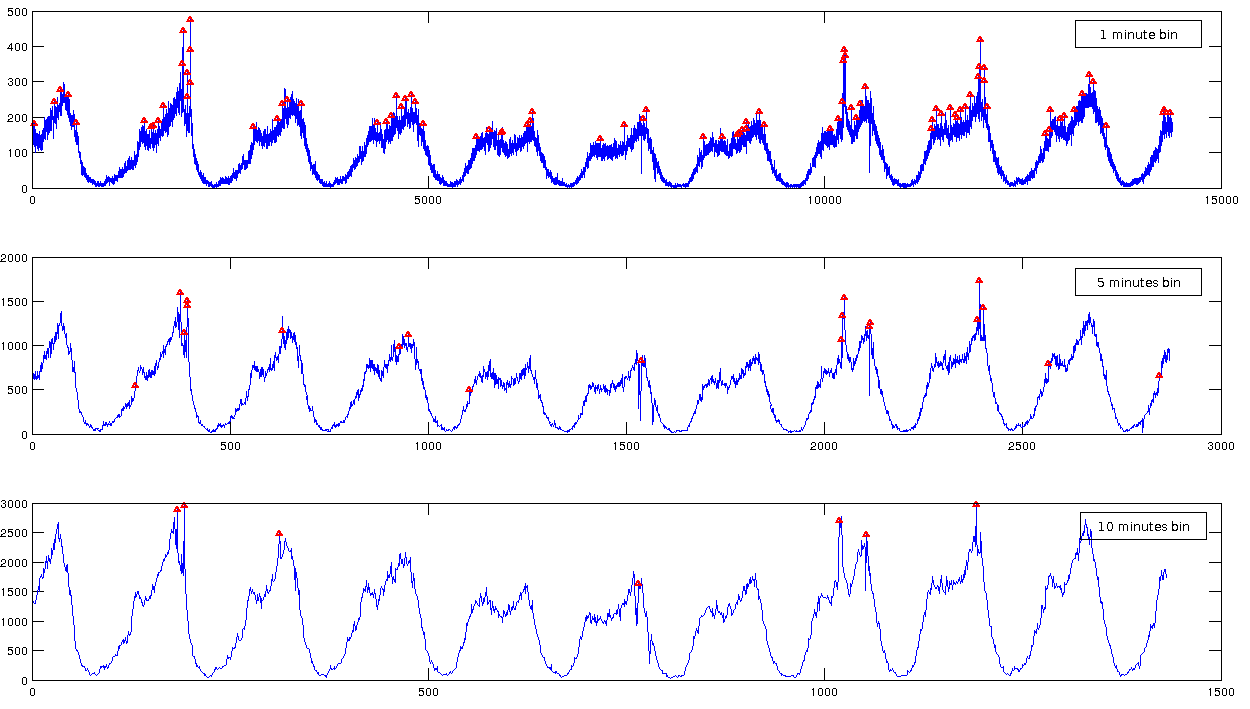}
\caption{Tweets time series on different bin's size and the detected outliers \label{figSaoPauloDiffTime}}
\end{figure}

We begin by examining the performance of the outliers' detection against the number of events occurred, unique, duplicated and missed events. Events occurred are events that happened in the real world and that were evaluated using the most frequent words in the messages of each bin matching with the result of a local newspaper's web search, using the time instance date as filter. We test the bin's size parameter for 1, 5 and 10 minutes, over the same period (Figure \ref{figSaoPauloDiffTime}), the precision rate score is presented with the mentioned metrics (Table \ref{tabDiffTime}). As bin size increases, it smooth local data likelihood making outliers the only values with significant difference. Otherwise, some not so substantial events occurred are missed.

\begin{table}[t]
\caption{Precision rate scores on different bin's size\label{tabDiffTime}}
\centering{
\begin{tabular}{|r|p{35pt}|p{48pt}|p{35pt}|p{50pt}|p{35pt}|p{36pt}|}
\hline
\textbf{Bin Size} & \textbf{Total Outliers} & \textbf{Detected Happened Events} & 
\textbf{Unique Events} & \textbf{Duplicate Detections} & \textbf{Missed Events} & \textbf{Precision Rate}\\
\hline
1 minute & 90 & 22 & 6 & 16 & 0 & 24.44\%\\
\hline
5 minutes & 20 & 12 & 4 & 8 & 2 & 60.00\%\\
\hline
10 minutes & 7 & 5 & 3 & 2 & 3 & 71.43\%\\
\hline
\end{tabular}}
\end{table}

\begin{figure}[t]
\centering
\includegraphics[width=1\textwidth]{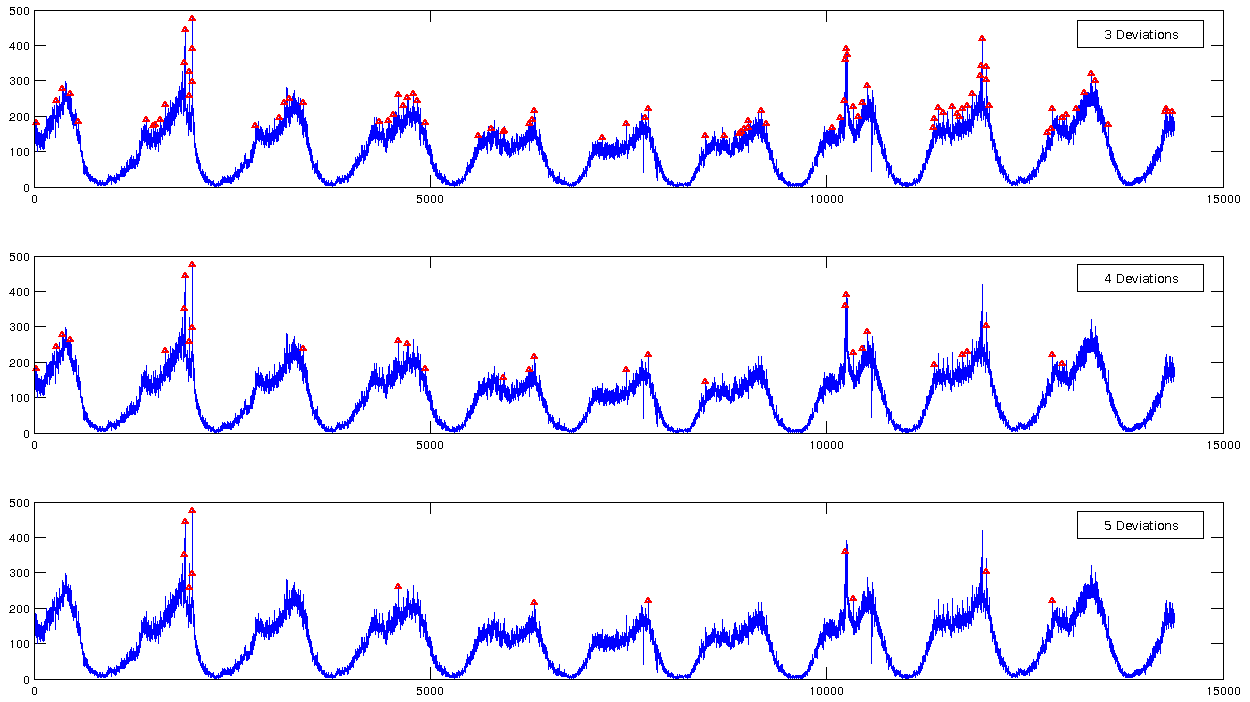}
\caption{Tweets time series on different deviation and the detected outliers \label{figSaoPauloDiffDev}}
\end{figure}

The next parameter evaluated, standard deviation, was tested with a time instance of 1 minute size and different values of deviations, i.e., 3, 4 and 5. Not surprisingly, the number of outliers detected decreased as the deviation increased (Figure \ref{figSaoPauloDiffDev}), but the change on the precision rate did not evolve like the previous experiment (Table \ref{tabDiffDev}). Our first assumption is that the 1-minute bin makes the time series rough and sensible to any minimum disturbances, making the deviation parameter tune incapable of getting better results. On the other hand, just increasing the bin's size will cause the loss of the real-time approach capability, as well as of some events. Therefore, a suggested approach is to combine the tuning of these parameters (a task that is reserved for future work).

\begin{table}[t]
\caption{Precision rate scores on different deviations\label{tabDiffDev}}
\centering{
\begin{tabular}{|p{48pt}|p{40pt}|p{50pt}|p{35pt}|p{48pt}|p{36pt}|p{36pt}|}
\hline
\textbf{Standard Deviations} & \textbf{Total Outliers} & \textbf{Detected Happened Events} & 
\textbf{Unique Events} & \textbf{Duplicate Detections} & \textbf{Missed Events} & \textbf{Precision Rate}\\
\hline
3 & 90 & 22 & 6 & 16 & 0 & 24.44\%\\
\hline
4 & 31 & 11 & 5 & 6 & 1 & 35.48\%\\
\hline
5 & 12 & 8 & 3 & 5 & 3 & 66.67\%\\
\hline
\end{tabular}}
\end{table}

In the task of evaluating the outliers with real-world events, the use of the most frequent terms allows us to understand the kinds of topics that trigger Twitter users to post significantly more messages than the usual. Firstly, we must understand that cultural aspects can influence social media services usage, so our findings consider, yet, only São Paulo's social behavior. All events occurred detected by our framework had televised coverage, but some with broad and other with local geographical interest (Table \ref{tabEvents}). This leads us to new perspectives of specializing event detection with only local relevance.

\begin{table}[b]
\caption{Events identified by the proposed approach\label{tabEvents}}
\centering{
\begin{tabular}{|p{160pt}|p{150pt}|p{57pt}|}
\hline
\textbf{Event Description} & \textbf{Terms} & \textbf{Geographical Interest}\\
\hline
Soccer match for Copa Libertadores in Venezuela & Corinthians, jogo, libertadores, gol, timão & Broad\\
\hline
National reality TV show & Yuri, fael, bbb, lider, ganhar & Broad\\
\hline
Soccer match on regional championship out the city & Corinthians, willian, douglas, gol, jogo & Broad\\
\hline
Riots at carnival vote counting & Gavi\~{o}es, carnaval, nota, fogo, apura\c{c}\~{a}o, escola & Local\\
\hline
Two soccer games in the regional championship out the city & Gol, jogo, bragantino, time, corinthians & Broad\\
\hline
Soccer match on regional championship in the city & Ganhar, vergonha, deus, palmeiras & Local\\
\hline
\end{tabular}}
\end{table}

\section{Conclusions}\label{sec:conclusion}

This paper presented a new method to discover events based on location over the Twitter stream, using time series analysis, and how this approach can lead to representative outliers with no need to previously select keywords, nor use clustering algorithms for geographic location grouping. This work provides the first step in a series of method to improve the detection of events with local relevance.

In future work, we will generate statistical measures of performance and compare our proposition with Lee's and Becker's method, and how those frameworks behave in a real-time environment, which can show how IGMN reuse benefits the performance. To do this comparison, we need to compute Lee's aggregation and dispersion metric, but other metrics with linear processing time can be built in order to consider the users' movement. To compute our method's precision and recall rate we intend to use human annotators and a news database to automate the events evaluation. A visualization system is suggested to provide more relevant information to the end user.



\bibliographystyle{jidm}
\bibliography{augusto}

\end{document}